\documentclass{ws-procs9x6}

\usepackage{bm,epsfig}

\begin{document}

\title{Pentaquarks in SU(3) Quark Model}

\author{Hungchong Kim}

\address{Department of Physics, Pohang University of Science and
Technology \\
Pohang 790-784, Korea\\
E-mail: hungchon@postech.ac.kr}

\author{Yongseok Oh}

\address{Department of Physics and Astronomy, University of Georgia \\
Athens, Georgia 30602, USA\\
E-mail: yoh@physast.uga.edu}
\maketitle

\abstracts{
Based on the flavor SU(3) symmetry,  we classify all the possible
pentaquark baryons made of four quarks and one antiquark.
In particular, we present possible multiplets of pentaquarks, their
interactions with mesons, and the mass relations within a multiplet.
We also study the pentaquark decays in the generalized OZI
rule.
}

\section{Introduction}

After the first report on the pentaquark
$\Theta^+(1540)$~\cite{LEPS03} and its subsequent
confirmation~\cite{theta:pos}, there has been a huge amount of
works studying pentaquark properties theoretically and experimentally.
Experimentally the subsequent observation of $\Xi^{--}(1862)$ by
NA49 Collaboration~\cite{NA49-03}
may suggest that $\Xi(1862)$ forms pentaquark antidecuplet with
$\Theta^+(1540)$ as anticipated by the soliton model~\cite{DPP97}.
Later the H1 Collaboration reported on the existence of anti-charmed
pentaquark state~\cite{H1-04}, which revives the interests in the heavy
pentaquark system~\cite{Lip87-GSR87,heavy,KLO04}.
However, the existence of pentaquark baryons is not fully
confirmed yet as some experiments report null results for
those states~\cite{theta:null,xi:null}.
A summary for the experimental situation and perspectives can be found,
e.g., in Refs.~\refcite{CLAS04a}.
Theoretically, many ideas have been put forward to study the
exotic pentaquark states in various approaches and
models~\cite{JM03,Jaffe04,LKK04}, but more detailed studies are required to
understand the properties and formation of pentaquark states.

As the pentaquark baryons may be produced in photon-hadron or
hadron-hadron reactions, it is important to understand their production
mechanisms and decay channels in order to confirm the existence of the
pentaquark states and to study their properties.
The present studies on the production reactions, however, are
limited by the lack
of experimental and phenomenological inputs on some
couplings~\cite{LK03,OKL03,DKST03-Roberts04,ZA03-YCJ03,NT03-OKL04}.
In particular, those studies do not include the contributions from the
intermediate pentaquark states in production mechanisms.
Therefore it is desirable to classify the pentaquarks based on the
flavor SU(3) symmetry albeit the experimental uncertainties for their
existence.
The general classification is useful not only for identifying
possible pentaquarks but also for providing selection rules for
their decays and the mass relations.

Of course, one can expect certain mixing
among the possible multiplets.
Indeed, many theoretical speculations suggest that the physical
pentaquark states would be mixtures of various
multiplets~\cite{JW03,CD04,EKP04}.
Thus it is necessary to construct the wavefunctions of pentaquark
baryons in terms of quark and antiquark for understanding the structure of
pentaquark states.
In this talk, we classify all
the pentaquarks in SU(3) quark model and obtain their SU(3) symmetric
interactions with other baryons.
Then mass relations among the pentaquark baryons will be presented.
In addition, we explore the special case when the antidecuplet-octet
ideal mixing is imposed to the pentaquark baryons.
The topics presented in this manuscript are discussed in more detail in
Refs.~\refcite{OKL03b,LKO04,OK04}.

\section{ General classification of pentaquarks}

The SU(3) flavor symmetry is a nice platform to construct
possible pentaquark multiplets.  Having four-quark and one
antiquark, the possible multiplets for the pentaquarks are
\begin{equation}
\bm{3} \otimes \bm{3} \otimes \bm{3} \otimes \bm{3} \otimes
\overline{\bm{3}} = \bm{35} \oplus (3)\bm{27} \oplus (2)\overline{\bm{10}}
\oplus (4) \bm{10} \oplus (8) \bm{8} \oplus (3)\bm{1}.
\label{group}
\end{equation}
Thus, we expect 91 different ground-state pentaquarks in total.
Of course, possible multiplets can be reduced under model assumptions.
We name all the pentaquarks based on hypercharge and isospin.
We denote the first subscript as the multiplet that the resonance
sits in and the second subscript as the isospin.
For $\Sigma$-like resonance, for example,
we have $\Sigma_{27,2}$
belonging to the 27-plet with isospin 2.
The superscript
will be reserved for the charge. Obvious indices will be suppressed
for simplicity.

\begin{figure}[ht]
%\epsfxsize=10cm   %width of figure - will enlarge/reduce the figures
%\epsfbox{fig3.eps}
%\figurebox{2cm}{3cm}{} %to have a box alone
\centerline{\epsfxsize=3.1in\epsfbox{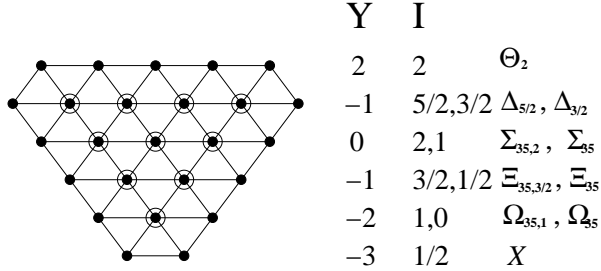}}
\caption{The weight diagram for the 35-plet. Our assignment
for the resonances are presented according to the specified
hypercharge and
isospin. In this diagram, the resonance $X$ is special by its
hypercharge and isospin.
 \label{fig1}}
\end{figure}

It is straightforward to construct the highest
multiplet from Eq.(\ref{group}). The weight diagram for the
35-plet and our assignment for the resonances are shown in
Fig.~\ref{fig1}.   The resonance $X$, which has hypercharge $-3$
and isospin $1/2$, is interesting for future search. By
eliminating the states at the corners in Fig.~\ref{fig1}, one can
generate the weight
diagram for the 27-plet.  Further successive elimination generates
decuplet, antidecuplet, octet and singlet.  All the resonances
other than the 35-plet are listed in Table~\ref{table1}.

%\begin{table}[t]
\begin{table}[ph]
\tbl{Pentaquark baryons except the 35-plet. For the 35-plet, see
Fig.~\ref{fig1}.}
%\centering
{\footnotesize
\begin{tabular}{c|cc|l} \hline\hline
multiplet & hypercharge & isospin & particle \\ \hline
$\bm{1}$ & $0$ & $0$ & $\Lambda_1^0$ \\ \hline
$\bm{8}$  & $1$ & $1/2$ & $N_{8}^{+},N_8^{0}$ \\
          & $0$ & $1$ & $\Sigma_8^{+},\Sigma_8^0,\Sigma_8^-$ \\
          & $0$ & $0$ & $\Lambda_8^{0}$ \\
          & $-1$ & $1/2$ & $\Xi_8^{0}, \Xi^-_{8}$ \\ \hline
$\bm{10}$ & $1$ & $3/2$ & $\Delta_{10}^{++},\Delta_{10}^+,\Delta_{10}^0,
                           \Delta_{10}^{-}$ \\
          & $0$ & $1$ & $\Sigma_{10}^{+},\Sigma_{10}^0, \Sigma_{10}^{-}$ \\
          & $-1$ & $1/2$ & $\Xi^{0}_{10},\Xi_{10}^-$ \\
          & $-2$ & $0$ & $\Omega_{10}^-$ \\ \hline
$\overline{\bm{10}}$ & $2$ & $0$ & $\Theta^{+}$ \\
                     & $1$ & $1/2$ & $N_{\overline{10}}^{+},
                                      N_{\overline{10}}^0$ \\
                     & $0$ & $1$ & $\Sigma_{\overline{10}}^{+},
\Sigma_{\overline{10}}^0,\Sigma_{\overline{10}}^-$ \\
                     & $-1$ & $3/2$ & $\Xi_{\overline{10},3/2}^{+}$,
     $\Xi_{\overline{10},3/2}^0$, $\Xi_{\overline{10},3/2}^-$,
     $\Xi_{\overline{10},3/2}^{--}$ \\ \hline
$\bm{27}$ & $2$ & $1$ & $\Theta_1^{++}, \Theta_1^+, \Theta_1^0$ \\
          & $1$ & $3/2$ & $\Delta_{27}^{++}$,  $\Delta_{27}^{+}$,
             $\Delta_{27}^{0}$,  $\Delta_{27}^{-}$ \\
          & $1$ & $1/2$ & $N_{27}^+$, $N_{27}^0$ \\
          & $0$ & $2$ & $\Sigma_{27,2}^{++}$,  $\Sigma_{27,2}^{+}$,
$\Sigma_{27,2}^0$, $\Sigma_{27,2}^-$, $\Sigma_{27,2}^{--}$\\
  & $0$ & $1$ & $\Sigma_{27}^+$, $\Sigma_{27}^0$, $\Sigma_{27}^-$ \\
          & $0$ & $0$ & $\Lambda_{27}^{0}$ \\
          & $-1$ & $3/2$ & $\Xi_{27,3/2}^+$, $\Xi_{27,3/2}^{0}$,
          $\Xi_{27,3/2}^-$, $\Xi_{27,3/2}^{--}$ \\
          & $-1$ & $1/2$ & $\Xi_{27}^{0}$, $\Xi_{27}^-$ \\
          & $-2$ & $1$ & $\Omega_{27,1}^0$, $\Omega_{27,1}^-$,
          $\Omega_{27,1}^{--}$ \\
\hline\hline
\end{tabular}\label{table1} }
\vspace*{-13pt}
\end{table}

\section{Tensor notation and SU(3) Lagrangians}

To get the SU(3) symmetric Lagrangian, it is useful to
represent all the resonances in the tensor notation.
In the tensor notation, the $(p,q)$ type of Young tableaux is represented
by the tensor, $T^{b_1 b_2, ...,b_q}_{a_1 a_2,...,a_p}$, which is
completely symmetric in upper and lower indices.
Also it is traceless on every pair of upper and lower indices.
In the tensor notation, all the pentaquark multiplets can be
represented by
\begin{eqnarray}
&& {\bf 1}: S\ , \quad  {\bf 8}: P^{i}_{j}\ ,
\quad {\bf 10} : D_{ijk}\ , \nonumber \\
&& {\bf \overline{10}} : T^{ijk} \ , \quad {\bf 27} : T^{ij}_{kl}\
, \quad {\bf 35} : T^{k}_{ijlm} \ .
\end{eqnarray}
Since we know how the quarks (and antiquarks) transform in SU(3),
one can easily construct transformation rules for the upper and
lower indices separately. Then one can assign each
resonance with a specific tensor or linear combination of
tensors within a multiplet~\cite{OK04}.
The SU(3) symmetric lagrangians can be constructed by
collecting all the possible contractions of upper and
lower indices. This is only way to form a SU(3) invariant
under SU(3) transformation. These interactions give selection rules that
can be used to search for specific pentaquarks.

For the interactions between pentaquark--baryon-octet--meson-octet, we obtain
\begin{eqnarray}
{\bf 1_5 - 8_3}  &:&  g_{\bm{1}\mbox{-}\bm{8}_3}^{}
\overline{S} \, B^i_k M^k_i
+ \mbox{(H.c.)} \nonumber \\
{\bf 8_5 - 8_3} &:& (d+f) \bar{P}_i^l B_k^i M_l^k +
(d-f) \bar{P}_i^l B_l^k M_k^i
+ \mbox{(H.c.)} \nonumber \\
{\bf 10_5 - 8_3} &:& g_{\bm{10}\mbox{-}\bm{8}_3}^{} \,
\epsilon_{ijk} \overline{D}^{jlm} B^i_l M^k_m + \mbox{(H.c.)} \nonumber \\
{\bf \overline{10}_5 - 8_3} &:&
g^{}_{\overline{\bm{10}}\mbox{-}\bm{8}_3} \,
\epsilon^{ilm}
\overline{T}_{ijk} B_l^j M_m^k + \mbox{(H.c.)} \nonumber \\
{\bf 27_5 - 8_3} &:& g_{\bm{27}\mbox{-}\bm{8}_3}^{} \,
\overline{T}^{kl}_{ij} B^i_k M^j_l + \mbox{(H.c.)}\ .
\end{eqnarray}
Here we note that the 35-plet can not couple to the baryon octet and meson
octet.  For $8_5 - 8_3$ case, there are two possible
ways to form fully contracted terms, which lead to the famous
$f$- and $d$-type interactions.  It is somewhat painful to write
down all the interactions in terms of the resonances that we have
identified above but it can be done~\cite{OK04}.

As a second set, we have pentaquark interactions with
baryon decuplet and meson octet.
They are
\begin{eqnarray}
{\bf 8_5 - 10_3} &:& g_{\bm{8}\mbox{-}\bm{10}_3}^{} \,
\epsilon^{ijk} \overline{P}^l_i D^{(3)}_{jlm} M^m_k
+ \mbox{(H.c.)} \nonumber \\
{\bf 10_5 - 10_3} &:& g_{\bm{10}\mbox{-}\bm{10}_3}^{} \,
\overline{D}^{jkl} D^{(3)}_{mkl} M^m_j
+ \mbox{(H.c.)} \nonumber \\
{\bf 27_5 - 10_3} &:& g_{\bm{27}\mbox{-}\bm{10}_3}^{} \,
\epsilon^{imn} \overline{T}_{ij}^{kl} D^{(3)}_{mkl} M^j_n
+ \mbox{(H.c.)}\nonumber \\
{\bf 35_5 - 10_3} &:& g_{\bm{35}\mbox{-}\bm{10}_3}^{} \,
\overline{T}^{ijkl}_a D_{ijk} M^a_l
+ \mbox{(H.c.)} \ .
\end{eqnarray}
The interactions of pentaquark--pentaquark--meson-octet can be constructed
similarly and they can be found in Ref.~\refcite{OK04}.

One remark is that the pentaquarks in the 35-plet can be measured
in decuplet-octet decay.  If $X$ in the 35-plet exists, it can be
measured in the unique decay mode
\begin{eqnarray}
X^- (X^{--}) \rightarrow \overline{K}^0 \Omega^- ( K^- \Omega^-)\
.
\end{eqnarray}
This decay mode is not affected by the mixing among the
multiplets.

\section{Mass relations}

To derive mass relations among pentaquarks, we note
first that QCD mass terms can be written by 
%\begin{eqnarray}
%M_{QCD}
%&=& \bar{q} \left( \begin{array}{ccc} m_u & 0 & 0 \\ 0 & m_d & 0 \\ 0 & 0
%& m_s \end{array} \right) q
%\stackrel{m_u = m_d} {\longrightarrow} a \bar{q}^i q_i + b\bar{q}^i Y^j_i q_j
%\nonumber \\
%{\rm where}&& Y= \frac{1}{3}\left( \begin{array}{ccc} 1 & 0 & 0 \\
%0 & 1 & 0 \\ 0 & 0
%& -2 \end{array} \right).
%\end{eqnarray}
\begin{eqnarray}
M_{QCD}
&=& \left ( \begin{array}{ccc} \bar{u} & \bar{d} & \bar{s} \end{array}
\right ) 
\left( \begin{array}{ccc} m_u & 0 & 0 \\ 0 & m_d & 0 \\ 0 & 0
& m_s \end{array} \right) 
\left ( \begin{array}{c} u \\ d \\ s \end{array} \right )
\stackrel{m_u = m_d} {\longrightarrow} a \bar{q}^i q_i + b\bar{q}^i Y^j_i q_j
\nonumber \\
{\rm where}&& Y= \frac{1}{3}\left( \begin{array}{ccc} 1 & 0 & 0 \\
0 & 1 & 0 \\ 0 & 0
& -2 \end{array} \right).
\end{eqnarray}
This gives a simple recipe to construct mass terms to leading
order in SU(3) breaking:
the mass term can be constructed by making fully contracted
terms including hypercharge matrix $Y$.  Within this
prescription, we obtain the following mass formulas
for each multiplet,
\begin{eqnarray}
{\bf 8}_5 &:& a \overline{P}^i_j P^j_i + b
\overline{P}^i_j Y^l_i P^j_l
    + c \overline{P}^i_j Y^j_l P^l_i
\nonumber \\
{\bf 10}_5 &:& a \overline{D}^{ijk} D_{ijk} + b \overline{D}^{ijk}
Y^l_k D_{ijl}
\nonumber \\
{\bf \overline{10}}_5 &:& a \overline{T}_{ijk} T^{ijk} +
b \overline{T}_{ijk} Y_l^k T^{ijl}
\nonumber \\
{\bf 27} &:& a \overline{T}^{ij}_{kl} T^{kl}_{ij} +
b \overline{T}^{ij}_{kl} Y^l_m T^{km}_{ij} +
c \overline{T}^{ij}_{kl} Y^m_j T^{kl}_{im}
\nonumber \\
{\bf 35} &:& a \overline{T}^{jklm}_i T^{i}_{jklm} +
b \overline{T}^{jklm}_i Y^n_j T^{i}_{nklm} +
c \overline{T}^{jklm}_i Y^i_n T^{n}_{jklm}
\ .
\end{eqnarray}
Note, the parameters $a,b,c$ in different multiplets should be different.
Here, for the octet, we obtain the pentaquark analog of
Gell-Mann--Okubo mass formula
\begin{eqnarray}
2(N_8 + \Xi_8) = 3 \Lambda_8 + \Sigma_8\ ,
\end{eqnarray}
equal-spacing rule for the decuplet and antidecuplet
\begin{eqnarray}
&&\Omega_{10} - \Xi_{10} = \Xi_{10} - \Sigma_{10} = \Sigma_{10} -
\Delta_{10} \nonumber \\
&&\Xi_{\overline{10},3/2} - \Sigma_{\overline{10}} =
\Sigma_{\overline{10}} - N_{\overline{10}} =
N_{\overline{10}} - \Theta \ .
\end{eqnarray}
We find additional mass relations for the 27-plet
\begin{eqnarray}
&&
3(\Sigma_{27} +\Theta_1) = 2(\Delta_{27} + 2 N_{27}), \quad
3(\Xi_{27,3/2} +2 \Theta_1) = 4\Delta_{27}+ 5 N_{27},
\nonumber \\
&& 3(\Xi_{27} +2 \Theta_1) =  \Delta_{27} + 8 N_{27}, \quad
3(\Omega_{27,1} +3 \Theta_1) = 2 (\Delta_{27} + 5 N_{27})\ ,
\end{eqnarray}
from which we derive the GMO type relation
\begin{eqnarray}
2(N_{27}+\Xi_{27}) = 3\Lambda_{27} +\Sigma_{27}\ ,
\end{eqnarray}
and two relations of equal-spacing-rule type
\begin{eqnarray}
\Omega_{27,1} - \Xi_{27,3/2} &=&
\Xi_{27,3/2} - \Sigma_{27,2},
\nonumber \\
\Sigma_{27,2} - \Delta_{27} &=&
\Delta_{27} - \Theta_1.
\end{eqnarray}
For the 35-plet, we similarly obtain the followings
\begin{eqnarray}
&& \Omega_{35} - \Xi_{35} = \Xi_{35} - \Sigma_{35} = \Sigma_{35} -
\Delta_{35} = \Delta_{35} - \Theta_2, \nonumber \\ && X -
\Omega_{35,1} = \Omega_{35,1} - \Xi_{35,3/2} = \Xi_{35,3/2} -
\Sigma_{35,2} = \Sigma_{35,2} - \Delta_{5/2}\ , \nonumber \\ &&
5(\Theta_2 +\Sigma_{35,2}) = 2 (2 \Delta_{5/2} + 3 \Delta_{35}),
\nonumber \\ && 5(\Sigma_{35} - \Sigma_{35,2}) = -4 (\Delta_{5/2}
- \Delta_{35})\ , \nonumber \\ && 5(\Xi_{35} - 2 \Sigma_{35,2}) =
-8 \Delta_{5/2} +3 \Delta_{35}\ , \nonumber \\ && 5(\Omega_{35} -
3 \Sigma_{35,2}) = -2 (6 \Delta_{5/2} - \Delta_{35})\ .
\end{eqnarray}

\section{Decay modes in the generalized OZI rule }

\begin{table}[ph]
\tbl{Couplings of the ideally mixed pentaquark baryon states.
The subscripts represent states with either purely light
quark-antiquark pairs or purely strange quark-antiquark pairs.  }
%\centering
%\begin{ruledtabular}
{\footnotesize
\begin{tabular}{cc|cc|cc|cc} \hline\hline
\multicolumn{2}{c}{$N^+_{\bar{q}q}$} &
\multicolumn{2}{c}{$N^+_{\bar{s}s}$} &
\multicolumn{2}{c}{$N^0_{\bar{q}q}$} &
\multicolumn{2}{c}{$N^0_{\bar{s}s}$}
\\ \hline
$\pi^+ n$ & $-\sqrt6$ & $K^+ \Lambda$ & $-\frac{3}{\sqrt2}$ &
$\pi^0 n$ & $\sqrt{3}$ &
$K^0 \Sigma^0 $ & $-\sqrt{\frac{3}{2}}$ \\
$\pi^0 p$ & $-\sqrt3$ & $K^+ \Sigma^0$ & $\sqrt{\frac{3}{2}}$ &
$\pi^- p$ & $-\sqrt6$
& $K^+ \Sigma^-$ & $\sqrt3$ \\
$\eta_{\bar{q}q}^{} p$ & $\sqrt3$ & $K^0 \Sigma^+$ & $\sqrt{3}$ &
$\eta_{\bar{q}q}^{}$ & $\sqrt3$&
$K^0 \Lambda$ & $-\frac{3}{\sqrt2}$ \\
& & $\eta_{\bar{s}s}^{} p$ & $-\sqrt3$ && & $\eta_{\bar{s}s}^{} n$ &
$-\sqrt3$ \\
\hline \multicolumn{2}{c}{$\Sigma_{\bar{q}q}^+$} &
\multicolumn{2}{c}{$\Sigma_{\bar{s}s}^+$} &
\multicolumn{2}{c}{$\Sigma^-_{\bar{q}q}$} &
\multicolumn{2}{c}{$\Sigma^-_{\bar{s}s}$}
\\ \hline
$\pi^+ \Sigma^0$ & $\sqrt{\frac{3}{2}}$ & $K^+ \Xi^0$ &
$\sqrt6$ & $\pi^-\Sigma^0$ & $-\sqrt{\frac{3}{2}}$ & $K^0 \Xi^-$ & $-\sqrt6$ \\
$\pi^0 \Sigma^+$ & $-\sqrt{\frac{3}{2}}$ & $\eta_{\bar{s}s}^{}
\Sigma^+$ & $-\sqrt6$
& $\pi^0 \Sigma^-$ & $\sqrt{\frac{3}{2}}$ & $\eta_{\bar{s}s}^{} \Sigma^-$
& $-\sqrt6$ \\
$\eta_{\bar{q}q}^{} \Sigma^+ $ & $\sqrt{\frac{3}{2}}$ & &
& $\eta_{\bar{q}q}^{} \Sigma^-$ & $\sqrt{\frac{3}{2}}$ & & \\
$\pi^+ \Lambda$ & $-\frac{3}{\sqrt{2}}$ & & & $\pi^- \Lambda$
& $-\frac{3}{\sqrt{2}}$ & & \\
$\bar{K}^0 p$ & $-\sqrt3$ & & & $K^- n$ & $-\sqrt3$ & & \\
\hline \multicolumn{2}{c}{$\Sigma_{\bar{q}q}^0$} &
\multicolumn{2}{c}{$\Sigma_{\bar{s}s}^0$} & \multicolumn{2}{c}{} &
\multicolumn{2}{c}{}
\\ \hline
$\pi^+ \Sigma^-$ & $-\sqrt{\frac{3}{2}}$ & $K^+ \Xi^-$ &
$-\sqrt3$ & & & \\
$\pi^- \Sigma^+$ & $\sqrt{\frac{3}{2}}$ & $K^0 \Xi^0$ & $-\sqrt3$
& & & & \\
$\eta_{\bar{q}q}^{} \Sigma^0 $ & $\sqrt{\frac{3}{2}}$
& $\eta_{\bar{s}s}^{} \Sigma^0$ & $-\sqrt6$ & & & & \\
$\pi^0 \Lambda$ & $-\frac{3}{\sqrt{2}}$ & & & & & & \\
$K^- p$ & $-\sqrt{\frac{3}{2}}$ & & & & & \\
$\bar{K}^0 n$ & $\sqrt{\frac{3}{2}}$ & & & & & \\
\hline\hline
\end{tabular}\label{table2}}
%\end{ruledtabular}
\vspace*{-13pt}
\end{table}

In the diquark-diquark-antiquark model of pentaquarks, Jaffe and Wilczek
advocated the ideal mixing of the antidecuplet with the octet~\cite{JW03}.
In this picture, the two diquarks form $\overline{6}_f$. By combining
with the antiquark of $\overline{3}_f$, one can form pentaquarks
belonging to the antidecuplet and octet,
$\overline {\bf 6} \otimes \overline {\bf 3} = \overline {\bf 10}
\oplus {\bf 8}$.
This multiplication in the tensor notation can be
written as
\begin{eqnarray}
S^{ij} \otimes \bar{q}^k = T^{ijk} \oplus S^{[ij,k]}.
\label{tensor1}
\end{eqnarray}
Obviously, the last part, being an octet representation, can be
replaced by a two-index field $P^j_i$ as
\begin{eqnarray}
S^{[ij,k]}=\epsilon^{ljk} P_l^i+\epsilon^{lik}P_l^j\ .
\label{tensor2}
\end{eqnarray}
The separation into the two terms in the right-hand side is
necessary to make it symmetric in $i$ and $j$. If one assumes that
the pentaquark decay goes through the fall-apart mechanism, the
index $k$ in Eq.~(\ref{tensor2}), the index for the antiquark,
should be contracted with the antiquark index of the meson field.
This is in fact equivalent to the generalized OZI rule where the
quark-connected diagram dominates over the quark-annihilated
diagram~\cite{LKO04} in the pentaquark decay. In this approach,
the interaction take the form
\begin{eqnarray}
\mathcal{L}_{8}= g_{8}^{} \epsilon^{ilm} \overline{S}_{[ij,k]}
B^j_l M_m^k + \mbox{(H.c.)}.
\label{interaction2}
\end{eqnarray}
Substituting Eq.~(\ref{tensor2}) into Eq.~(\ref{interaction2}), one has
\begin{eqnarray}
\mathcal{L}_{8} = 2 g_8^{} \overline{P}^m_i B^i_l M^l_m + g_8
\overline{P}^m_i M^i_l B^l_m+ \mbox{(H.c.)}.
\end{eqnarray}
Comparison with the standard expression for the octet baryon
interactions leads to $f=1/2$ and $d=3/2$.
Therefore, the OZI rule makes a special choice on the $f/d$ ratio as
$f/d=1/3$~\cite{CD04,LKO04}.

In Table~\ref{table2}, we present the decay modes of
pentaquarks in the ideal mixing.  The $\bar{s} s$ component has
been separated in the quark wavefunctions for pentaquark baryons,
normal baryons and normal mesons. The pentaquarks that do not
suffer from the ideal mixing will have the same decay modes
as presented in the earlier sections.

\section{Summary}

We have classified all the pentaquark baryons in the flavor SU(3). 
The tenor
method has been facilitated in constructing their interactions
with normal baryons and mesons.  We have also presented the mass
relations for the pentaquarks which take into account the SU(3)
breaking to leading order. This will help to identify not only
exotic baryons but also crypto-exotic states.  Finally we have
discussed the decay modes in the generalized OZI rule, which turns
out be equivalent to the ideal mixing or fall-apart mechanism.


\begin{thebibliography}{10}

\bibitem{LEPS03}
LEPS Collaboration, T.~Nakano {\em et~al.\/},
  Phys. Rev. Lett. {\bf 91}, 012002 (2003).
%%CITATION = HEP-EX 0301020;%%

\bibitem{theta:pos}
DIANA Collaboration, V.~V.~Barmin {\em et~al.\/},
  Phys. At. Nucl. {\bf 66}, 1715 (2003);
%%CITATION = HEP-EX 0304040;%%
CLAS Collaboration, S.~Stepanyan {\em et~al.\/},
  Phys. Rev. Lett. {\bf 91}, 252001 (2003);
%%CITATION = HEP-EX 0307018;%%
SAPHIR Collaboration, J.~Barth {\em et~al.\/},
  Phys. Lett. B {\bf 572}, 127 (2003);
%%CITATION = HEP-EX 0307083;%%
CLAS Collaboration, V.~Kubarovsky {\em et~al.\/},
  Phys. Rev. Lett. {\bf 92}, 032001 (2004);
%%CITATION = HEP-EX 0311046;%%
A.~E. Asratyan, A.~G. Dolgolenko, and M.~A. Kubantsev,
  Phys. At. Nucl. {\bf 67}, 682 (2004);
%%CITATION = HEP-EX 0309042;%%
HERMES Collaboration, A.~Airapetian {\em et~al.\/},
  Phys. Lett. B {\bf 585}, 213 (2004);
%%CITATION = HEP-EX 0312044;%%
SVD Collaboration, A.~Aleev {\em et~al.\/},
  hep-ex/0401024;
%%CITATION = HEP-EX 0401024;%%
COSY-TOF Collaboration, M.~Abdel-Bary {\em et~al.\/},
  Phys. Lett. B {\bf 595}, 127 (2004);
%%CITATION = HEP-EX 0403011;%%
P.~Zh. Aslanyan, V.~N. Emelyanenko, and G.~G. Rikhkvitzkaya,
  hep-ex/0403044;
%%CITATION = HEP-EX 0403044;%%
ZEUS Collaboration, S.~Chekanov {\em et~al.\/},
  Phys. Lett. B {\bf 591}, 7 (2004);
%%CITATION = HEP-EX 0403051;%%
Yu. A.~Troyan {\em et~al.\/},
  hep-ex/0404003;
%%CITATION = HEP-EX 0404003;%%
S.~V. Chekanov for the ZEUS Collaboration,
  hep-ex/0404007.
%%CITATION = HEP-EX 0404007;%%

\bibitem{NA49-03}
NA49 Collaboration, C.~Alt {\em et~al.\/},
  Phys. Rev. Lett. {\bf 92}, 042003 (2004).
%%CITATION = HEP-EX 0310014;%%

\bibitem{DPP97}
D.~Diakonov, V.~Petrov, and M.~Polyakov,
  Z. Phys. A {\bf 359}, 305 (1997).
%%CITATION = HEP-PH 9703373;%%

\bibitem{H1-04}
H1 Collaboration, A.~Aktas {\em et~al.\/},
  hep-ex/0403017.
%%CITATION = HEP-EX 0403017;%%

\bibitem{Lip87-GSR87}
H.~J. Lipkin, Phys. Lett. B {\bf 195}, 484 (1987);
%%CITATION = PHLTA,B195,484;%%
C.~Gignoux, B.~Silvestre-Brac, and J.~M. Richard,
  Phys. Lett. B {\bf 193}, 323 (1987).
%%CITATION = PHLTA,B193,323;%%

\bibitem{heavy}
Y.~Oh, B.-Y. Park, and D.-P. Min,
  Phys. Lett. B {\bf 331}, 362 (1994);
%%CITATION = PHLTA,B331,362;%%
  Phys. Rev. D {\bf 50}, 3350 (1994);
%%CITATION = PHRVA,D50,3350;%%
Y.~Oh and B.-Y. Park,
  Phys. Rev. D {\bf 51}, 5016 (1995);
%%CITATION = PHRVA,D51,5016;%%
M.~Genovese, J.-M. Richard, Fl. Stancu, and S.~Pepin,
  Phys. Lett. B {\bf 425}, 171 (1998);
%%CITATION = HEP-PH 9712452;%%
Fl. Stancu,
  Phys. Rev. D {\bf 58}, 111501 (1998).
%%CITATION = HEP-PH 9803442;%%

\bibitem{KLO04}
H.~Kim, S.~H. Lee, and Y.~Oh,
  Phys. Lett. B {\bf 595}, 293 (2004).
%%CITATION = HEP-PH 0404170;%%

\bibitem{theta:null}
BES Collaboration, M.~Ablikim {\em et~al.\/},
  Phys. Rev. D {\bf 70}, 012004 (2004);
%%CITATION = HEP-EX 0402012;%%
K.~T. Kn{\"o}pfle {\em et~al.\/} for the HERA-B Collaboration,
  J. Phys. G {\bf 30}, S1363 (2004);
%%CITATION = HEP-EX 0403020;%%
HERA-B Collaboration, I.~Abt {\em et~al.\/},
  hep-ex/0408048;
%%CITATION = HEP-EX 0408048;%%
C.~Pinkenburg,
  nucl-ex/0404001;
%%CITATION = NUCL-EX 0404001;%%
SPHINX Collaboration, \mbox{Yu}. M.~Antipov {\em et~al.\/},
  hep-ex/0407026;
%%CITATION = HEP-EX 0407026;%%
BABAR Collaboration, B.~Aubert {\em et~al.\/},
  hep-ex/0408037;
%%CITATION = HEP-EX 0408037;%%
  hep-ex/0408064.
%%CITATION = HEP-EX 0408064;%%

\bibitem{xi:null}
H.~G. Fischer and S.~Wenig,
  hep-ex/0401014;
%%CITATION = HEP-EX 0401014;%%
STAR Collaboration, S.~Kabana {\em et~al.\/},
  hep-ex/0406032;
%%CITATION = HEP-EX 0406032;%%
J.~Pochodzalla,
  hep-ex/0406077;
%%CITATION = HEP-EX 0406077;%%
WA89 Collaboration, M.~I.~Adamovich {\em et~al.\/},
  Phys. Rev. C {\bf 70}, 022201 (2004);
%%CITATION = HEP-PH 0405042;%%
I.~V. Gorelov for the CDF Collaboration,
  hep-ex/0408025.
%%CITATION = HEP-EX 0408025;%%

\bibitem{CLAS04a}
V.~D. Burkert {\em et~al.\/},
  nucl-ex/0408019;
%%CITATION = NUCL-EX 0408019;%%
V.~Kubarovsky and P.~Stoler,
  hep-ex/0409025;
%%CITATION = HEP-EX 0409025;%%
K.~Hicks,
  hep-ph/0408001.
%%CITATION = HEP-PH 0408001;%%

\bibitem{JM03}
B.~K. Jennings and K.~Maltman, Phys. Rev. D {\bf 69}, 094020 (2004).
%%CITATION = HEP-PH 0308286;%%

\bibitem{Jaffe04}
R.~L. Jaffe, hep-ph/0409065.
%%CITATION = HEP-PH 0409065;%%

\bibitem{LKK04}
S.~H.~Lee, H.~Kim and Y.~Kwon,
  Phys.\ Lett.\ B {\bf 609}, 252 (2005)
%%CITATION = HEP-PH 0411104;%%

\bibitem{LK03}
W.~Liu and C.~M. Ko,
  Phys. Rev. C {\bf 68}, 045203 (2003);
%%CITATION = NUCL-TH 0308034;%%
  Nucl. Phys. A {\bf 741}, 215 (2004);
%%CITATION = NUCL-TH 0309023;%%
W.~Liu {\em et al.\/},
  Phys. Rev. C {\bf 69}, 025202 (2004).
%%CITATION = NUCL-TH 0310087;%%

\bibitem{OKL03}
Y.~Oh, H.~Kim, and S.~H. Lee,
  Phys. Rev. D {\bf 69}, 014009 (2004);
%%CITATION = HEP-PH 0310019;%%
  Phys. Rev. D {\bf 69}, 074016 (2004);
%%CITATION = HEP-PH 0311054;%%
Y.~Oh, H.~Kim and S.~H.~Lee,
  Nucl.\ Phys.\ A {\bf 745}, 129 (2004).
%%CITATION = HEP-PH 0312229;%%

\bibitem{DKST03-Roberts04}
A.~R. Dzierba {\em et al.\/},
  Phys. Rev. D {\bf 69}, 051901 (2004);
%%CITATION = HEP-PH 0311125;%%
W.~Roberts,
  nucl-th/0408034.
%%CITATION = NUCL-TH 0408034;%%

\bibitem{ZA03-YCJ03}
Q.~Zhao and J.~S. Al-Khalili,
  Phys. Lett. B {\bf 585}, 91 (2004);
  {\bf 596}, 317(E) (2004);
%%CITATION = HEP-PH 0312348;%%
B.-G. Yu, T.-K. Choi, and C.-R. Ji,
  nucl-th/0312075.
%%CITATION = NUCL-TH 0312075;%%

\bibitem{NT03-OKL04}
K.~Nakayama and K.~Tsushima,
  Phys. Lett. B {\bf 583}, 269 (2004);
%%CITATION = HEP-PH 0311112;%%
K.~Nakayama and W.~G. Love,
  Phys. Rev. C {\bf 70}, 012201 (2004);
%%CITATION = HEP-PH 0404011;%%
Y.~Oh, K.~Nakayama and T.~S.~Lee,
 hep-ph/0412363.
%%CITATION = HEP-PH 0412363;%%


\bibitem{JW03}
R.~Jaffe and F.~Wilczek,
  Phys. Rev. Lett. {\bf 91}, 232003 (2003).
%%CITATION = HEP-PH 0307341;%%

\bibitem{CD04}
F.~E. Close and J.~J. Dudek,
  Phys. Lett. B {\bf 586}, 75 (2004).
%%CITATION = HEP-PH 0401192;%%

\bibitem{EKP04}
J.~Ellis, M.~Karliner, and M.~Prasza{\l}owicz,
  JHEP {\bf 05}, 002 (2004).
%%CITATION = HEP-PH 0401127;%%

\bibitem{OKL03b}
Y.~Oh, H.~Kim, and S.~H. Lee,
  Phys. Rev. D {\bf 69}, 094009 (2004).
%%CITATION = HEP-PH 0310117;%%

\bibitem{LKO04}
  S.~H.~Lee, H.~Kim and Y.~Oh,
  J.\ Korean Phys.\ Soc.\  {\bf 46}, 774 (2005).
%%CITATION = HEP-PH 0402135;%%

\bibitem{OK04}
  Y.~Oh and H.~Kim,
  Phys.\ Rev.\ D {\bf 70}, 094022 (2004).
  %%CITATION = HEP-PH 0405010;%%

\bibitem{PS04}
S.~Pakvasa and M.~Suzuki,
  Phys. Rev. D {\bf 70}, 036002 (2004).
%%CITATION = HEP-PH 0402079;%%

\bibitem{GS04}
S.~M. Golbeck and M.~A. Savrov,
  hep-ph/0406060.
%%CITATION = HEP-PH 0406060;%%


\end{thebibliography}
\end{document}